\documentclass[sigconf, nonacm]{acmart}

\usepackage{pifont}
\usepackage{nicefrac}
\AtBeginDocument{%
  }

\usepackage{subcaption}
\usepackage{graphicx}
\begin{document}

\title{Extracting Alpha from Financial Analyst Networks}


\author{Dragos Gorduza}
\affiliation{%
  \institution{University of Oxford}
  \city{Oxford}
  \country{United Kingdom}}
\email{dragos.gorduza@st-annes.ox.ac.uk}
\author{Yaxuan Kong} %
\affiliation{%
 \institution{University of Oxford}
  \city{Oxford}
  \country{United Kingdom}
  \email{dragos.gorduza@st-annes.ox.ac.uk}
}
\author{Xiaowen Dong}
\affiliation{%
 \institution{University of Oxford}
  \city{Oxford}
  \country{United Kingdom}
  \email{dragos.gorduza@st-annes.ox.ac.uk}
  }
\author{Stefan Zohren}
\affiliation{%
  \institution{University of Oxford}
  \city{Oxford}
  \country{United Kingdom}
  \email{dragos.gorduza@st-annes.ox.ac.uk}
  }

\renewcommand{\shortauthors}{D.Gorduza, Y.Kong, X.Dong, and S.Zohren}

\begin{abstract}
We investigate the effectiveness of a momentum trading signal based on the coverage network of financial analysts.  
This signal builds on the key information-brokerage role financial sell-side analysts play in modern stock markets. The baskets of stocks covered by each analyst can be used to construct a network between firms whose edge weights represent the number of analysts jointly covering both firms. Although the link between financial analysts coverage and co-movement of firms' stock prices has been investigated in the literature, little effort has been made to systematically learn the most effective combination of signals from firms covered jointly by analysts in order to benefit from any spillover effect. 
To fill this gap, we build a trading strategy which leverages the analyst coverage network using a graph attention network. More specifically, our model learns to aggregate information from individual firm features and signals from neighbouring firms in a node-level forecasting task. 
We develop a portfolio based on those predictions which we demonstrate to exhibit an annualized returns of 29.44\% and a Sharpe ratio of 4.06 substantially outperforming market baselines and existing graph machine learning based frameworks. We further investigate the performance and robustness of this strategy through extensive empirical analysis.  Our paper represents one of the first attempts in using graph machine learning to extract actionable knowledge from the analyst coverage network for practical financial applications. 

\end{abstract}





\maketitle

\section{Introduction}

Financial analysts play an important information-processing role in modern financial markets.   They provide signals about firm health, and their predictions about future firm outcomes have traditionally been regarded as a shorthand for stock pickers to easily keep their finger on the `pulse' of the market. However, simultaneous coverage of two firms by one or more analysts is also shown to explain higher levels of return correlation between those firms' returns \citep{ali2020shared,jiang2023shared,yi2023pair}. This suggests that beyond providing information, sell-side analyst coverage also has an effect on investor attention which we propose to leverage in order to build a trading strategy. 

Indeed, as investors follow specific analyst reports, they tend to focus their finite attention on specific baskets of firms covered by sell-side analysts \citep{israelsen2016does}. 
Limited investor attention has been identified as a source of investment opportunities \citep{ali2020shared} as information overload leads to delayed recognition of the impact that news affecting one company should have on other economically related companies. This slow information diffusion can result in predictable lead-lag effects between multiple firm's returns \citep{cohen2008economic}. 
By acting as conduits for investor attention, `analyst coverage networks' have been used to explain uneven patterns of investor attention between firms in the market \citep{israelsen2016does,jiang2023shared}. Analyst coverage networks refer to networks where nodes are firms and the weighted edges between two firms denote the number of analysts covering both firms in a certain time frame such as the last year or the last quarter \citep{ali2020shared}. For instance, if Analyst A covers both Google and Apple, a link with weight 1 is established between these two firms in the network. If Analyst B also covers Google and Apple, the link between the two companies will have a weight 2.  
These networks are interesting to financial research beyond just directing investor attention. Analyst coverage networks are shown to also identify fundamental economic linkages between firms, as analysts tend to cover related firms \citep{ali2020shared}. High analyst coverage offers strong predictive power for financial linkages \citep{gomes2023analyst} as well as changes in firms’ fundamental information \citep{lee2017uncovering,bekkerman2023effect,hameed2015information}. This motivates our interest in leveraging the analyst coverage network to build a profitable momentum-spillover strategy. 

This increased co-movement between stocks linked to analyst coverage has led to initial interest \citep{Oyeniyi2020} in creating a `model-free' strategy to profit from this effect. This approach is easy to interpret and implement as it involves simple weighted averages of neighbour returns as indicators of future profitability. 
The trading signal then takes the form of averaging momentum of a firms' 1-hop neighbours in the network as an indicator for future performance. However, the neighbourhood momentum indicator - as we investigate in this work - may be too weak on its own and fail to take full advantage of the nonlinear and multi-hop relationships between a firm and its wider neighbourhood. Moreover, existing approaches mainly focus on the pre-defined network structure, and hence lack the flexibility of adapting the strength of the relationships between firms, especially given node-level information coming from the stock price evolution of firms in the portfolio. 

Lead-lag models developed in research and industry meanwhile have begun including machine learning in uncovering firm-level momentum to build trading indicators \citep{baz2015dissecting}. Inspired by this, we propose to leverage the additional flexibility allowed by graph machine learning models to 1) include both firm level and network level data 
and 2) adjust and adapt in real-time the strength of relationships from the initial analyst coverage network, 
in order to build a trading signals for out-of-sample returns. This allows us to profit from the rich information incorporated in the analyst network which uncovers relationships between stocks not traditionally captured by industry or correlation-based linkages. In addition, we demonstrate the potential of a multi-layer graph attention network (GAT) \citep{casanova2018graph} in capturing the non-linear and multi-hop relationships in the analyst network in order to build an improved trading strategy in real time. 


Our research contributes to the existing literature in several ways. First, to the best of our knowledge, our work represents one of the first attempts in using graph machine learning techniques to model the analyst coverage network and predict stock returns.
Second, we demonstrate the superiority of our GAT-based approach over traditional aggregation methods, highlighting the importance of capturing complex, non-linear relationships between firms in the analyst coverage network. Third, we show the benefits of incorporating higher-frequency firm-level data into the model, enabling it to adapt more effectively to changing market conditions, thereby generating more robust trading strategies.

The outline of this paper is as follows. We first discuss the relevant literature which led to the different components of our approach. Subsequently, we formulate the trading-strategy problem as a graph machine learning question and present the model itself as well as the methodology we built for evaluating the model against a suite of industry benchmarks and ablation studies. Lastly, we present and discuss the experimental results before concluding and presenting the potential avenues for future research.





\section{Literature review}
The concept of analyst coverage networks and their potential to explain various momentum spillover effects in the stock market has captured significant attention in recent literature.

Several studies have investigated the economic and fundamental linkages between companies to better understand and capitalize on the lead-lag effect in investment strategies. The effect is evident among companies within the same industry \citep{hou2007industry, moskowitz1999industries}, firms offering similar products \citep{cohen2012complicated, hoberg2018text}, entities connected through supplier-customer chains \citep{cohen2008economic, menzly2010market}, firms sharing common technological innovations \citep{bekkerman2023effect, lee2019technological}, those located in the same geographic region \citep{chen2022far, parsons2020geographic}, companies with overlapping institutional ownership \citep{gao2017institutional}, and those with the same strategic alliances \citep{cao2016alliances}. 

In addition to the relationships above, Ali and Hirshleifer \citep{ali2020shared} introduce shared analyst coverage as an additional overarching approach to determine firm relatedness. They specifically highlight its potential to explain various previously established cross-asset momentum effects such as industry momentum, geographic momentum or customer momentum. They argue that analyst linkages are particularly adept at uncovering fundamental relationships between companies, more so than other methods for identifying firm linkages such as industry networks and correlations. Investors that overlook the analyst coverage network may be underestimating fundamental channels of shock transmission between firms thus explaining the observed momentum spillover effect between strongly connected neighbours on the analyst coverage network \citep{Oyeniyi2020,israelsen2016does}. 
Ali and Hirshleifer \citep{ali2020shared} also argue that shared analyst coverage could quantify the strength of company relationships more accurately than simple binary variables or sector groupings, and address the challenges faced by other methods such as difficulty in accessing complete supplier information for a given company \citep{cohen2008economic}.

Recent papers on shared analyst coverage have further explored its potential to unify momentum spillover effects and predict returns. Key studies include Gomes et al. \citep{gomes2023analyst}, who investigated the role of analyst coverage networks in corporate financial policies. Yi and Guo \citep{yi2022common} provided evidence from China on how common analyst links predict returns, while Jiang et al. \citep{jiang2023shared} explored connected-firm momentum spillover in China. Oyeniyi et al. \citep{Oyeniyi2020} discussed profiting from sell-side analysts’ coverage networks, Israelsen \citep{israelsen2016does} 
suggested that investors make correlated information processing errors which can be tracked using the analyst coverage network and help explain the link between it and excess comovement, and Martens and Sextroh \citep{martens2021analyst} examined interfirm information spillovers due to overlapping analyst coverage. 

While the existing literature has made significant progress in understanding the role of analyst coverage networks in explaining momentum spillover effects, there remain several gaps that our research aims to address. Previous studies have primarily focused on using simple aggregation methods, such as weighted averages of a firm’s direct neighbours’ momentum, to create lead-lag portfolios \citep{ali2020shared}. However, these approaches may not fully capture the complex and non-linear relationships between firms in the analyst coverage network, limiting their ability to predict future stock returns accurately. Moreover, existing methods lack the flexibility to adapt the strength of the aggregation coefficients based on higher-frequency information from the firms in the portfolio, hindering the strategies’ ability to respond to rapidly changing market conditions and firm-specific events.

Our research aims to address these gaps by leveraging the power of graph machine learning techniques, specifically graph attention networks (GATs) \citep{casanova2018graph}, to develop a more sophisticated and adaptive approach to modeling the analyst coverage network and predicting stock returns. The GAT's attention based architecture allows our model to assign different importance weights to each firm’s neighbors in the analyst coverage network, thus making its' representation of firm to firm interactions more flexible than the static weights used in model-free approaches to building portfolios from the original analyst coverage network \citep{Oyeniyi2020}.

\section{Methods}
\subsection{Data}

\subsubsection{Data sources}
We investigate a dataset of stock prices extracted from the CRSPR/COMPUSTAT datasource hosted by WRDS covering the 2006-2022 period\footnote{Compustat prices are available at : https://wrds-www.wharton.upenn.edu/pages/get-data/center-research-security-prices-crsp/annual-update/crspcompustat-merged/security-daily/}. The dataset is composed of 495 firms of the SNP500 accross a diverse array of industries. We combine this dataset with information from Institutional Brokers' Estimate System  (IBES) analyst estimates from 2006 to 2022\footnote{IBES coverages are available at : https://wrds-www.wharton.upenn.edu/pages/about/data-vendors/vendor-partner-ibes/}. These estimates are recorded as they are produced by each analyst covering firms in our sample over the 2006-2022 period. These forward looking predictions are made at yearly quarterly or monthly horizons. 

\subsubsection{Momentum indicators from stock prices}
\label{sec:momentum}
We calculate $r_{i, t-\Delta t}$, the log-return over $\Delta$ time periods of firm $i$ as described in Equation \ref{eq:logreturns}. We select 5  $\Delta$ to represent several return horizons $\Delta \in [1, 21, 63, 126, 252]$ corresponding to the returns $r_{i, t-\Delta t}$ over the past 1, 21, 63, 126 and 252 days respectively:
\begin{equation}
   r_{i, t-\Delta t} = \log( \frac{p_{i, t}}{p_{i, t-\Delta}})
\label{eq:logreturns}
\end{equation}
With $p_{i, t}$ and $p_{i, t-\Delta}$ the price of security $i$ at times $t$ and $t-\Delta$ respectively. Those 5 return horizons serve as 5 common measurements for firm information as described in \citep{baz2015dissecting}.

We define 3 additional firm-level indicators following \cite{baz2015dissecting} $s_{i, t}(S,L)$ in Equation \ref{eq:macd} based on varying short (S) and long (L) time scales belonging to $(S,L) \in {(8, 24),(16, 48),(32, 96)}$. 




\begin{equation}
 s_{i, t}(S,L) = \frac{\frac{m(i, t, S)-m(i, t, L)}{std(r_{i, t-63 t})}}{{std(\frac{m(i, t, S)-m(i, t, L)}{std(r_{i, t-63})})}}
\label{eq:macd}
\end{equation}
Here $m(i,t,S) = \gamma * p_{i, t} + (1-\gamma) * m(i, t-1, S)$  is the exponential weighted moving average of the price of asset i at time t with a scaling factor $\gamma$ = $\frac{1}{S}$ and $std(r_{i, t-63 t})$ is the standard deviation of prices over the past 63 days. We group these 8 indicators for each firm $i$ into a vector $\vec{x}_{i, t}$ $\in$ $(1,8)$. With $\vec{x}_{i, t}$ =  $[r_{i, t-1t},r_{i, t-5t}, ...,  s_{i, t}(32,96)]$ of length 8. Combining each $\vec{x}_{i, t}$ into a matrix with all firms, we obtain the feature matrix $\mathbf{X}_t$ $\in$ $(N,8)$. 




\subsubsection{Network extraction from IBES ratings}
\label{sec:analyst-projection}
\begin{figure}[h!]
    \centering
    \includegraphics[width=\linewidth]{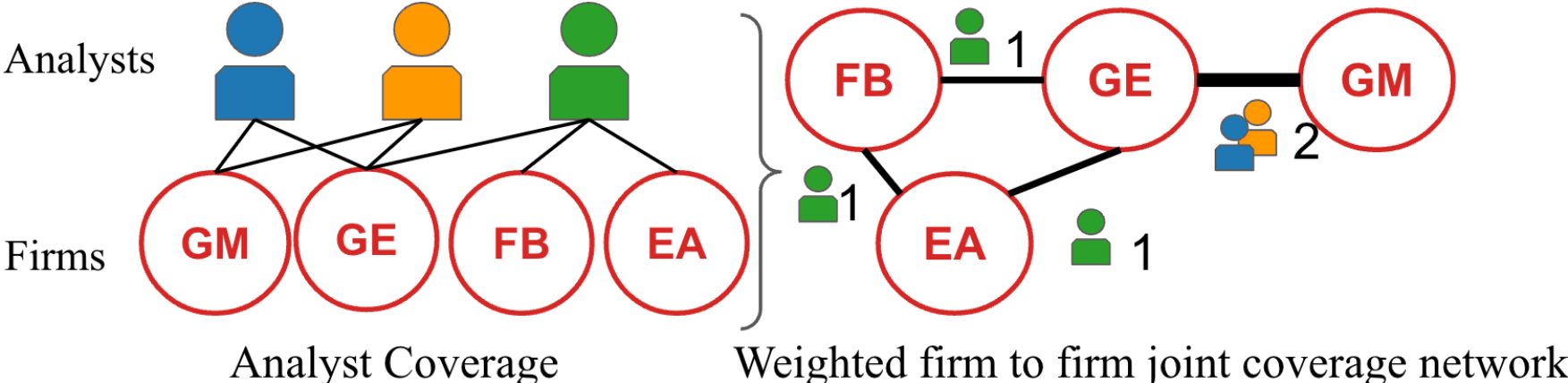}
    \caption{Building the analyst network}
    \label{fig:analyst_projection}
\end{figure}

We represent the overlapping coverage portfolios of the analysts in our sample as an undirected network. Each trading day in our sample, we record the IBES estimates made by each analyst for all the 495 firms in our sample over the past 252 days. This creates an evolving bipartite analyst-firm coverage record which we transform into an evolving firm-to-firm network by counting the number of analysts that cover every pair of firms as described in Figure \ref{fig:analyst_projection}. 
The nodes of the proposed analyst coverage network represent all the firms in our sample. Meanwhile an edge between two nodes represents the number of analysts that cover both firms within a given look back window of 252 days. This approach to defining analyst coverage is common in the literature studying the statistical properties of analysts' choices \citep{ali2020shared}. This allows us to create at each time step an adjacency matrix $\mathbf{A}_t$ that captures the topological information contained in the analyst coverage network on day $t$. 


\subsection{Problem formulation }
\label{sec:problem-formulation}
We propose to frame the task of constructing a portfolio of stocks 
as a classification task. 
We define the categorical target variable $y_{i, t+21}$ for each firm as the out-of-sample excessive returns compared to the average return of all firms in the market which takes as value either 1 (overperformance) or 0 (underperformance) 
as defined in Equation \ref{eq:y-target}:
\begin{equation}
y_{i, t+21} = 
\begin{cases}
    1 & \text{if } r_{i, t+21-t} > \frac{\sum_{i}^{N}{r_{i, t+21-t}}}{N}\\
    0              & \text{otherwise}
\end{cases}
\label{eq:y-target}
\end{equation}

Following \citep{Oyeniyi2020}, we select a monthly target as the statistical inquiries into the analyst matrix as we assume the monthly frequency is a reasonable timeframe for the effect of analyst coverage on investors to manifest itself.

We stack each of the target variables $y_{i, t+21}$ into a target vector $\mathbf{Y}_{t+21}$. 
We are aiming to define a function 
that produces the best $\mathbf{\hat{Y}}_{t+21}$, the out-of-sample forecast of over and under-performance. This forecast takes the form of a predicted probability of belonging to class 1 or 0.


Following \citep{roberts2023network}, we then transform  $\mathbf{\hat{Y}}_{t+21}$ into an investment strategy by buying the stocks corresponding to the entries in $\mathbf{\hat{Y}}_{t+21}$ with the 25\% highest predicted probability to over-perform. Similarly, we sell the stocks corresponding to the entries in $\mathbf{\hat{Y}}_{t+21}$ with the 25\% highest predicted probability to under-perform.  


\subsection{Proposed methodology}
\label{subsec:prop-methodology}
For a given trading day $t$ in our sample, we have a feature matrix $\mathbf{X}_t$, an adjacency matrix $\mathbf{A}_t$ and a target vector $\mathbf{Y}_{t+21}$. We use these in order to build the graph $G = {\mathbf{A}_t,\mathbf{X}_t}$ and train a GAT  that learns the mapping between network information and the target following the form $GAT(\mathbf{A}_t,\mathbf{X}_t)=\mathbf{Y}_{t+21}$. This triplet $\mathbf{A}_t$,$\mathbf{X}_t$ and $\mathbf{Y}_{t+21}$ defines one sample, subsection \ref{sec:training-strategy} describes how we combine these samples to form our training-validation-testing sets. 

The GAT layer takes as input a matrix of node features $\mathbf{X}_t = [\vec{x}_{1 t},\vec{x}_{2, t},..., \vec{x}_{N t}]$ where $N$ is the number of nodes and $\vec{x}_{1, t}$ is a vector of dimension $D$ where $D$ is the number of features in each input node. In our setup, as discussed in Section \ref{sec:momentum}, the number of dimensions is 8. We describe the mechanism for transforming each node  feature vector $\vec{x}_{1, t}$ to their updated value $\vec{x}'_{1, t}$ after applying each GAT layer.
It uses the attention function described in Equation \ref{eq:attention-function} which calculates the attention score between two vectors using a shared attentional mechanism $a : \mathbb{R}^D \times \mathbb{R}^D \rightarrow \mathbb{R}$:
\begin{equation}
    e_{ij, t} = a(\mathbf{W}\vec{x}_{i, t}\mathbf{W}\vec{x}_{j, t})
\label{eq:attention-function}
\end{equation}

The GAT also normalizes the attention scores using the softmax function to make them comparable between layers. 
\begin{equation}
    \alpha_{ij, t} = \text{softmax}(e_{ij, t}) = \frac{e_{ij, t}}{\sum_{k \in N(i)}{e_{ik, t}}}
\end{equation}
Having acquired the attention scores $\alpha$, the GAT layer applies a parametrization using a weight matrix $\mathbf{W}$ and a  non-linear transformation with the $ReLU$ function in order to obtain the updated feature representation $\vec{x}'_i$ as described in Equation \ref{eq:node-update}. 
\begin{equation}
    \vec{x}'_{i, t} = \text{ReLU}(\sum_{j \in N(i)}{\alpha_{ij, t} \mathbf{W} \vec{x}_{j, t}})
\label{eq:node-update}
\end{equation}

%

The model we present is set up to use an arbitrary number of GAT layers followed by a linear layer with learnable weight matrix $\mathbf{W}_{\text{linear}}$ as described in Equation \ref{eq:prediction}.

\begin{equation}
     \hat{Y}_{t+21} = ReLU(\mathbf{X}'\mathbf{W}_{\text{linear}})
\label{eq:prediction}
\end{equation}

\begin{figure}[h!]
  
  \includegraphics[width=\linewidth]{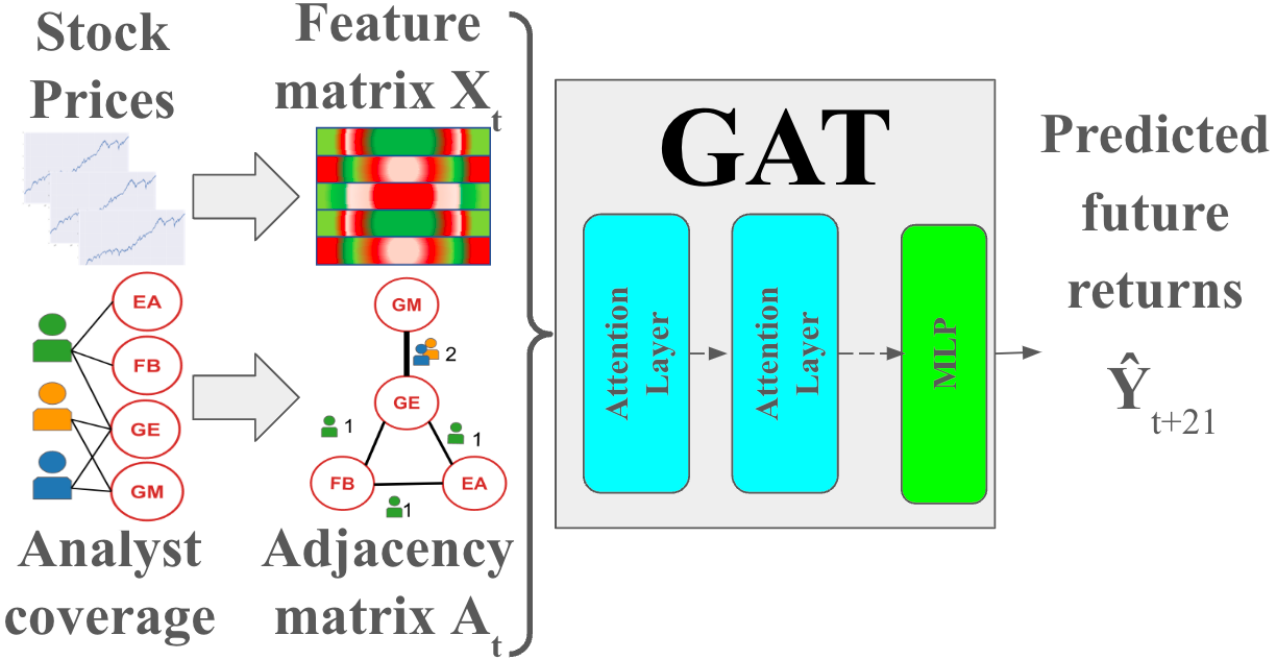}
  \caption{Pipeline of the proposed GAT trading model}
  \label{fig:pipeline-gat}
\end{figure}

We describe the overall pipeline of our model in Figure \ref{fig:pipeline-gat}. 

\subsection{Training strategy}
\label{sec:training-strategy}
We split our 17 year dataset into 204 1-month trading periods, at the end of each period, we retrain and validate our model. In order to obtain training, validation and testing sets we group multiple triplets $\mathbf{A}_t$,$\mathbf{X}_t$ and $\mathbf{Y}_{t+21}$ together. We group the first 10 samples in each trading period (from t=0 to t=9) to form our training set, we group the following 10 samples (from t=10 to t=20)  to form our validation set and lastly, we test our model on the 21st sample corresponding to t=21.  

For validation, we perform hyperparameter tuning using grid search with the following settings: the learning rate $\in$ \{1e-2,1e-3,1e-4\}, the layer size $\in$ \{64,128\}, the number of layers $\in$ \{1,2\}, the weight decay regularization $\in$ \{1e-4,1e-5,1e-6\}, and Attention heads $\in$ \{2,8\}. 





\subsection{Comparative baseline models}
We also consider several candidate approaches against which to compare the performance of our model. These all leverage either the feature matrix $\mathbf{X}_t$ or the network information $\mathbf{A}_t$ and serve as alternative formulations present in the literature for how to build a trading signal. These are summarized in Table \ref{tab:models} : 
\begin{enumerate}
    \item Market Long Only: buying all the stocks in the market with equal weights
    
    \item MACD Momentum: averaging the momentum indicators defined in Equation \ref{eq:macd} to use as a trading indicator.
    \item Analyst Matrix: following \cite{Oyeniyi2020}, averaging the momentum of the 1-hop neighbours on the analyst coverage matrix. 
    \item Neural Network (NN): Using a 2-layer feed-forward neural network to predict $\hat{Y}_{t+21}$
\end{enumerate}

%


\begin{table}[h!]
    \centering
    \caption{Comparison of features between our proposed model and benchmarks}
        \begin{tabular}{cccc}
            \toprule
            Model Name & uses $\mathbf{X}_t$ & uses $\mathbf{A}_t$ & Learning \\
            \midrule
            Long Only & \ding{56} & \ding{56} & \ding{56} \\
            MACD Momentum Averaging & \ding{52} & \ding{56} & \ding{56} \\
            Analyst Matrix & \ding{52} & \ding{56} & \ding{52} \\
            Neural Network & \ding{56} & \ding{52} & \ding{56} \\
            \midrule
            Ours (\text{GAT}$_\text{analysts}$) & \textbf{\ding{52}} & \textbf{\ding{52}} & \textbf{\ding{52}} \\
            \bottomrule
        \end{tabular}
    \label{tab:models}
\end{table}

\subsection{Ablation studies}
\label{subsec:ablation}
We also perform a series of ablation studies on the performance of our graph attention based model in different setups to understand what drives its performance and better grasp what features add value. To do this we take the basic setup described in \ref{subsec:prop-methodology} and replace different components of the basic \text{GAT}$_\text{analysts}$ model:
\begin{itemize}
    \item $GCN$: A graph convolutional Network \citep{kipf2016semi} learning model that does not use attention for propagating information.
    \item \text{GAT}$_\text{1\_layer}$ : A GAT model with only 1-layer instead of the 2 of our initial setup.
    \item \text{GAT}$_\text{corr}$ : A GAT model which uses as neighbourhood information a correlation matrix which we transform into an adjacency matrix by eliminating edges with correlation inferior to the 90-th percentile.
    \item \text{GAT}$_\text{industries}$ : A GAT model which uses as neighbourhood information a the GICS industrial classifications with firms being connected if and only if they are in the same industry
    \item \text{GAT}$_\text{del\_edge}$ : A GAT model which uses the original analyst network from which 60\% were randomly removed.
\end{itemize}

\subsection{Evalutation metrics}
We evaluate these strategies by calculating several features displayed by the returns of their corresponding portfolios. Those are:
\begin{itemize}
    \item Returns: the annualized average gross percent returns of strategy 
    \item  Volatility: their annualized average standard deviation of percent returns over the time period denoted as Vol.
    \item Sharpe Ratio: the ratio of the average annualized returns minus the risk-free interest rate divided by the standard deviation, a measure of risk-adjusted returns of the portfolio. 
    \item Maximum Drawdown (MD): the maximum peak-to-trough span reached by the portfolio
    \item Maximum Drawdown duration (MDD): the maximum number of consecutive periods the portfolio was in drawdown expressed as a percentage of the entire number of periods
\end{itemize}
Moreover, we also calculate the cumulative log returns of each strategy by adding up the log returns at each trading period. This gives us a final measure of how well the strategy performed in gross returns.
\section{Results}
This result section will first present the main predictive comparisons in Section \ref{sec:predictive-perf}, then discuss the ablation studies in Section \ref{sec:ablation-study}. We analyse the correlation between the returns and the market in Section \ref{sec:corr-analysis}. We perform a turnover analysis of the different strategies in Section \ref{subsec:turnover}. Lastly, we present discuss the behaviour of the attention weights in the model in Section \ref{subsec:attention}.
\label{sec:results}

    

\subsection{Comparison of predictive performances}
\label{sec:predictive-perf}
\begin{table}[!h]
    \centering
    \caption{Performance metrics of different portfolios}

    \begin{tabular}{cccccc}
            \toprule
            & \begin{tabular}{@{}c@{}}Returns \\ (\%)\end{tabular} & \begin{tabular}{@{}c@{}}Vol. \\ (\%)\end{tabular} & Sharpe  & \begin{tabular}{@{}c@{}}MD \\ (\%)\end{tabular} & \begin{tabular}{@{}c@{}}MDD \\ (\%)\end{tabular} \\
            \midrule
            Market & 6.89 & 11.88 & 0.411  & -39.4 & 21.0 \\
            Analyst Matrix & 1.83 & 8.58 & 0.069  & -22.8 & 51.0 \\
            MACD & 9.56 & 11.46 & 0.672  & -35.3 & 19.0 \\
            NN & 15.44 & 8.32 & 1.753  & -6.4 & 4.0 \\
            \midrule
            \begin{tabular}{@{}c@{}}GAT $_\text{analysts}$\end{tabular} & \textbf{29.44} & \textbf{7.07} & \textbf{4.069}  & \textbf{-6.0} & \textbf{1.0} \\
            \bottomrule
        \end{tabular}
    \label{tab:returns}
\end{table}

Table \ref{tab:returns} presents the results of the different strategies against the financial tests we use to evaluate performance. The market long only shows a Sharpe ratio of 0.411 and returns of 6.89\% displaying relatively worse performances. It also displays the longest maximum drawdown with -39.4\% of value lost at the trough. The Analyst Matrix strategy consisting of a weighted average of neighbour momentum is the worst performing strategy in both returns (1.83\%) and risk adjusted returns (annualized Sharpe ratio of 0.069). It also performs the worst in terms of maximum drawdown duration with 51\% of the trading backtest period (corresponding to 103 months). The MACD strategy performs better than the Analyst Matrix and the market long-only strategy with a Sharpe ratio of 0.672, and it shows a maximum drawdown of -35\% and a maximum drawdown duration of 39 trading periods (21\%). The Neural Network outperforms all the previously introduced strategies with a Sharpe ratio of 1.753 (more than double that of the next best MACD) and a much lower maximum drawdown of -6.42\%. The Neural network strategy displays a shorter MDD than the MACD with 4.0\% compared to 19\%. Lastly, the GAT$_\text{analysts}$ method we introduce displays higher log-returns (at 29.44\% annualized returns) and a Sharpe ratio of 4.069, more than double the previous best Sharpe ratio of 1.753 for the NN strategy. The GAT$_\text{analysts}$ also features lower drawdowns (-6\%) and shorter MDD than any other competing strategy with 1\% of the trading periods corresponding to 2 months of trading time being spent in continuous drawdown. The GAT$_\text{analysts}$ portfolio, moreover, displays lower volatility than every other strategy with 7\% against 8.32\% for the NN, the next lowest volatility strategy.

\begin{figure}[h!]
  
  \includegraphics[width=\linewidth]{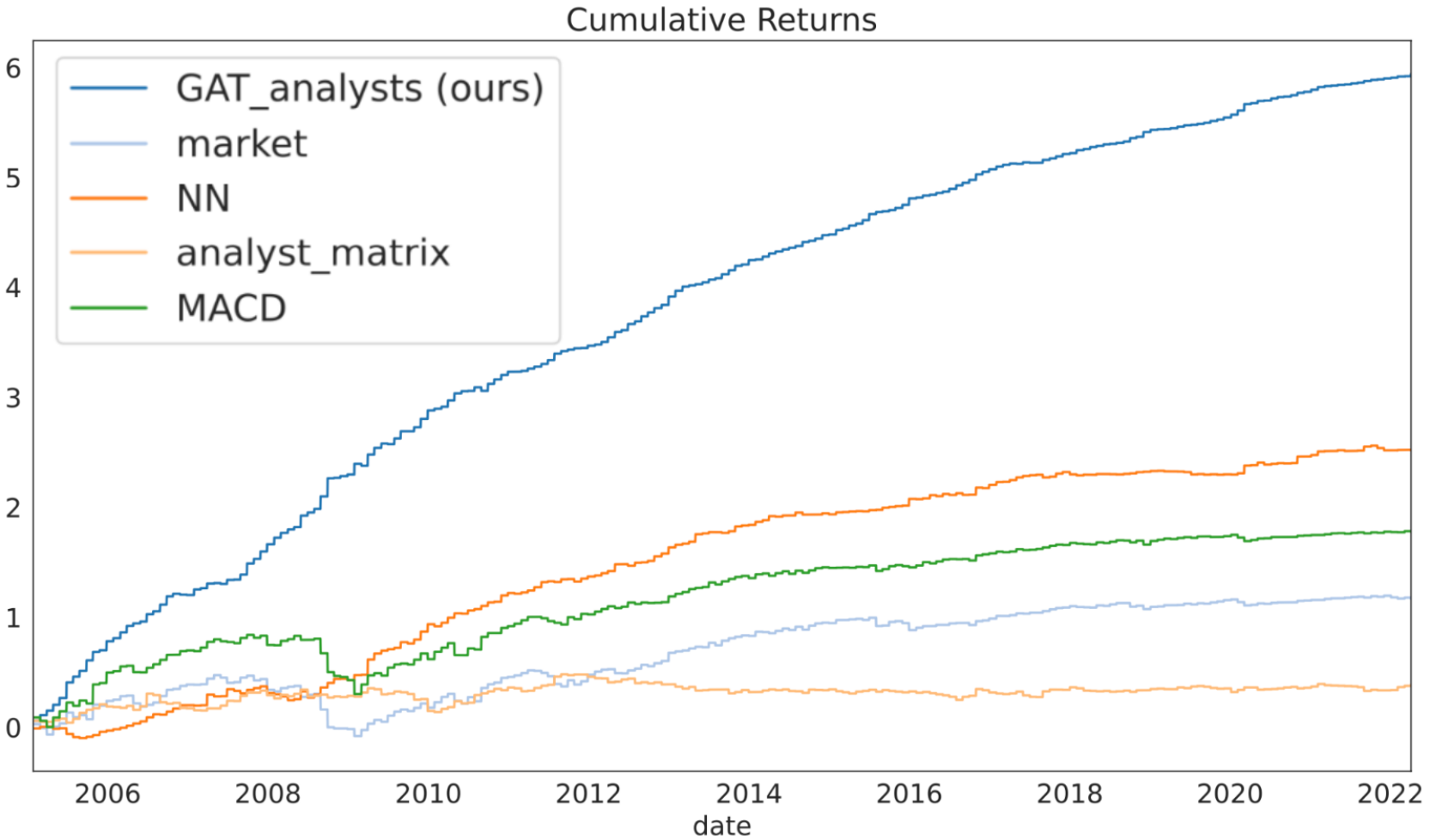}
  \caption{Cumulative returns}
  \label{fig:cumsum_returns_restricted}
\end{figure}%



The superiority of the GAT$_\text{analysts}$ model compared to its competitors can also be observed in Figure \ref{fig:cumsum_returns_restricted} which presents the cumulative log-returns. The GAT$_\text{analysts}$ strategy performs the best amongst all the presented approaches with a cumulative log-return of 5 over the 18 year period. The Neural network is the second best performing strategy throughout the evaluation period. It lags the market at the start of the period and quickly outperforms it and all the other approaches. This temporal evolution of the returns also serves to highlight the relative superiority of the GAT$_\text{analysts}$ after the 2008 financial crisis,  suggesting that it was able to identify promising lead-lag clusters despite the prevailing financial perturbation. Moreover, it also shows that the node and network information combined in the GAT$_\text{analysts}$ strongly outperforms the simple network aggregation strategy based on the Analyst matrix, which confirms our starting hypotesis that analyst coverage network topology and firm momentum features can be learnt jointly to extract alpha. 







\subsection{Ablation studies}
\label{sec:ablation-study}

    

\begin{table}[!h]
    \centering
    \caption{Ablation study}
        \begin{tabular}{cccccc}
            \toprule
            & \begin{tabular}{@{}c@{}}Returns \\ (\%)\end{tabular} & \begin{tabular}{@{}c@{}}Vol. \\ (\%)\end{tabular} & Sharpe  & \begin{tabular}{@{}c@{}}MD \\ (\%)\end{tabular} & \begin{tabular}{@{}c@{}}MDD \\ (\%)\end{tabular} \\
            \midrule
            \begin{tabular}{@{}c@{}}GAT $_\text{analysts}$\end{tabular} & 29.44 & \textbf{7.07} & \textbf{4.069}  & \textbf{-6.0} & \textbf{1.0} \\
            \begin{tabular}{@{}c@{}}GAT $_\text{corr}$\end{tabular} & \textbf{33.81} & 9.10 & 3.757  & -4.1 & 2.4 \\
            \begin{tabular}{@{}c@{}}GAT $_\text{industries}$\end{tabular} & 19.21 & 8.26 & 2.250 & -8.1 & 2.9  \\
            \begin{tabular}{@{}c@{}}GAT $_\text{del\_edge}$\end{tabular} & 19.57 & 8.26 & 2.265  & -10.3 & 5.3 \\
            \begin{tabular}{@{}c@{}}GAT $_\text{1\_layer}$\end{tabular} & 28.54 & 8.02 & 3.417 & \textbf{-3.6} & 1.5  \\
            GCN & 17.67 & 7.73 & 2.205 & -3.9 & 2.4  \\
            \bottomrule
        \end{tabular}
    \label{tab:returns-ablation}
\end{table}

%
%
%
%
%

Table \ref{tab:returns-ablation} presents the performance of the basic GAT framework exposed to different sources of network information instead of the analyst matrix 
We can observe that the $\text{GAT}_\text{analysts}$ outperforms all the alternative representations of firm to firm relationships with a Sharpe ratio of 4.069 compared with 3.757 for the \text{GAT}$_\text{corr}$ and less than 2.25 and 2.26 for the industry and edge\_delete versions. The $\text{GAT}_\text{analysts}$ displays slightly lower returns than the \text{GAT}$_\text{corr}$ (29.44\% vs 33.81\%), however the basic approach also displays lower volatility than any of the other ablations. The $\text{GAT}_\text{analysts}$ model which only uses the analyst matrix also displays a lower Maximum Drawdown duration (MDD) with only 1.0\% of the trading period being spent in drawdown, which is 50\% less than the next longest at 1.5\% of the \text{GAT}$_\text{1\_layer}$. However, we can observe that the peak-to-trough maximum drawdown of the \text{GAT}$_\text{corr}$ is less (-4.1\%) than the maximum drawdown of the $\text{GAT}_\text{analysts}$ (-6.0\%). The table above suggest that the information content of the alternative network formulations such as correlations, industries and edge-deletion do not allow the GAT to generate better risk-adjusted forecasts. The \text{GAT}$_\text{corr}$ displays slightly higher returns compared to the $\text{GAT}\_\text{analysts}$, however the higher volatility incurred by the correlation suggests that the analyst matrix helps the \text{GAT}$_\text{analysts}$ model to select firms with slightly lower return volatility, leading to better risk-adjusted returns. 

\begin{table}[!h]
    \centering
    \caption{Improvements from different model setups}
    
        \begin{tabular}{ccccc}
            \toprule
            \begin{tabular}{@{}c@{}} Model \\ Name \end{tabular} & \begin{tabular}{@{}c@{}}Cum. \\ returns\end{tabular} & \begin{tabular}{@{}c@{}}Message \\ Passing\end{tabular} & Adjacency & \begin{tabular}{@{}c@{}}\# of \\ Layers \end{tabular} \\
            \midrule
            \text{GAT}$_\text{analysts}$ & \textbf{5.9} & Attention & Analyst & 2 \\
            \text{GAT}$_\text{corr}$ & 5.4 & Attention & Correlation & 2 \\
            \text{GAT}$_\text{1\_layer}$ & 4.7 & Attention & Analyst & 1 \\
            \text{GAT}$_\text{industries}$ & 4.7 & Attention & Industries & 2 \\
            \text{GAT}$_\text{del\_edge}$ & 3.6 & Attention & Perturbed & 2 \\
            GCN & 3.0 & Convolution & Analyst & 2 \\
            Market & 1.2 & - & - & - \\
            \bottomrule
        \end{tabular}
    
    \label{tab:deltas}
\end{table}

Table \ref{tab:deltas} summarizes how much improvement we obtain in terms of cumulative returns for different message passing, adjacency information and number of layers for the models proposed in Section \ref{subsec:ablation} as ablation studies. Cumulative returns can be interpreted as the ability of a model to extract a trading signal from the node and network information.  
We can observe that the largest improvement
is achieved by introducing attention instead of graph convolutions for message passing. This significant improvement corresponds to a near doubling of cumulative returns over the trading period (+96\% between \text{GAT}$_\text{analysts}$ and GCN). In addition to this, we observe that the going from a one layer model (\text{GAT}$_\text{1\_layer}$) to a two layer model (\text{GAT}$_\text{analysts}$) led to an a 25\% increase in returns
Meanwhile having the complete analyst matrix instead of the randomly perturbed one leads to 63\% more returns. Replacing the industrial-GICS network of firms with the analyst matrix leads to a 25\% improvement in cumulative log returns. Lastly, replacing the correlation-based adjacency matrix with the analyst matrix nets a comparatively smaller but still meaningful 9\% increase in cumulative returns. These results comfort the initial hypothesis that the analyst matrix contains useful information for the building of a portfolio. This improvement can be attributed to the structural information present in the adjacency as removing edges from the analyst matrix and replacing the matrix with other firm-to-firm networks leads to strongly degraded results. 
Another useful observation can be drawn from the single vs multi-hop setup which suggests that it is not only the single-hop neighbourhood aggregation that brings value as suggested in \cite{Oyeniyi2020,ali2020shared}. Rather, the proper leveraging of the complex and informative relationships present in the analyst matrix requires more complex models able to aggregate information from wider neighbourhoods. Combining information from 2-hop neighbours would allow the GAT model to update firm representations to include   This could be the result of the analyst matrix helping to uncover latent relationships which are hard to detect otherwise, such as alignment of economic and actuarial practices amongst covered firms and the higher probability for an analyst to follow firms that use similar technological tools \cite{martens2021analyst}.    

\begin{table}[h!]
    \centering
    \caption{Topology comparisons between different networks}
        \begin{tabular}{cccc}
            \toprule
            Name & Jaccard & Diameter & Transitivity \\
            \midrule
            Industry & 1.0 & 1.0 & 1.0 \\
            Correlation & 0.34 & 5.4 & 0.66 \\
            Analyst & 0.98 & 11.29 & 0.67 \\
            \bottomrule
        \end{tabular}
    \label{tab:topology}
\end{table}

Table \ref{tab:topology} presents a comparison between the three network topologies considered in the ablation study. The Jaccard similarity (percentage of edges in common), Diameter (longest path between two nodes) and Transitivity (fraction of all possible triangles in the graph) are calculated at each time t for each graph. They are then averaged to produce on measure for each graph type and metric. The transitivity comparison shows that the correlation and analyst networks are less clustered than the industry network. That is expected as the industry network is a set of fully connected components. Moreover, the analyst and correlation networks both display similar levels of transitivity at 0.67 and 0.66 respectively, which signifies that over 60\% of open triangles are connected. A high transitivity suggests both the analyst and correlation networks are strongly clustered.
The network diameter in the Diameter column  describes the maximum distance between two nodes in the graph. Both the analyst and correlation matrices have a higher diameter (11 on average for the analyst network and 6 on average for the correlation network) compared to the industry network. A network with a higher diameter can help in reducing the likelihood of over smoothing in GAT models which may explain the better performance of the \text{GAT}$_\text{analysts}$. Moreover, a higher diameter suggests that the analyst coverage connects firms along `longer' chains. As discussed in \citep{Oyeniyi2020, ali2020shared, jiang2023shared}, these longer chains closely represent fundamental real-life links between firms which tend to be missed by the correlation matrix. This helps explains the added value of the analyst network : it captures different fundamental relationships \citep{cohen2008economic}. 
The jaccard index of the different networks presented the Jaccard column show that the analyst matrix remains consistently self-similar (over 90\%) as compared to the correlation matrix. The structural properties of the industry network will not evolve as the structure of the industry classifications which explains the high jaccard similarity of 1.0 (meaning the industry network stays on average constant through time). Meanwhile the correlation network displays a much lower jaccard index of 34\% implying it evolves faster period-to-period than the analyst matrix. A more stable network period to period can help with training a graph machine learning model and contributes to the better performance of the \text{GAT}$_\text{analysts}$ strategy. Existing literature \citep{pacreau2021graph} performs quantitative comparisons between industry and correlation based models as these are more commonly used to build graph-based trading strategies. However, few compare these networks with the analyst matrix as it is less often used as a building block of trading strategies. We show that these differences are profound and can serve to explain the different performance of a graph machine learning tool on each of these firm to firm networks. 

\subsection{Return correlation analysis}
\label{sec:corr-analysis}
\begin{figure}[h!]
\centering
\begin{subfigure}[t]{0.46\linewidth}
    \centering
    \includegraphics[width=1\linewidth]{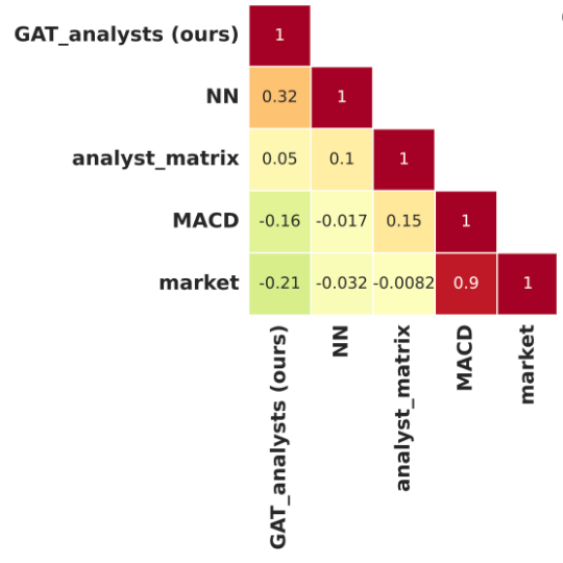}
    \caption{Benchmark methods}
    \label{fig:return_corr_restricted}
\end{subfigure}
\begin{subfigure}[t]{0.54\linewidth}
    \centering
    \includegraphics[width=1\linewidth]{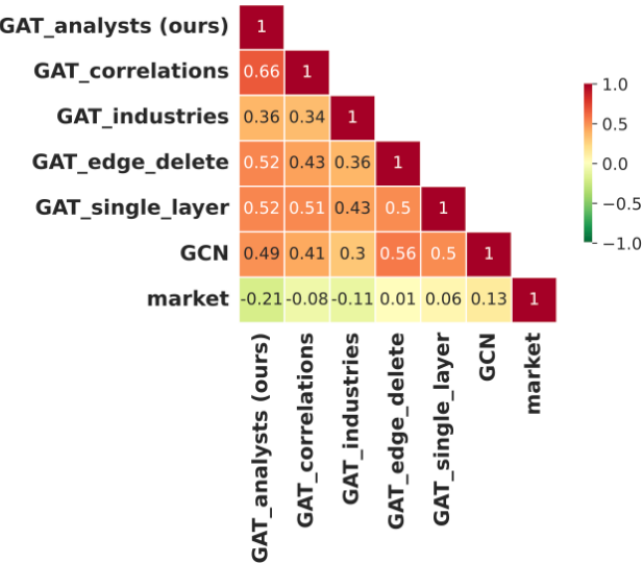}
    \caption{Ablation study}
    \label{fig:correlations_ablation}
 \end{subfigure}
 \caption{Correlation matrices of studied signals}
\end{figure}
Figure \ref{fig:return_corr_restricted} displays the correlations of returns between the different evaluated trading strategies. A high correlation coefficient in red indicates that the two models' returns follow similar movements. Low correlations are generally sought by investors as it allows for diversification. All else being equal investors can limit the concentration of their portfolio and increase their returns by investing in the least correlated signals.  
The GAT model appears most anti-correlated with the market with a -0.21 coefficient. The different models' returns appear uncorrelated with each other aside from a slightly higher return correlation between the $\text{GAT}_\text{analysts}$ and the NN of 0.32. These plots suggest that the returns of $\text{GAT}_\text{analysts}$ follow a pattern that responds to different market signals compared to the other competing models. This supports the hypothesis that the analyst network is a useful and informative prior with which to diversify a trading strategy as the signal it produces is different from the other existing ones. 
Figure \ref{fig:correlations_ablation} represents the correlation of the returns for strategies based on \text{GAT}$_\text{analysts}$ and all of the different ablations under investigation. The correlation of all of these approaches with the market are also presented. The original \text{GAT}$_\text{analysts}$ displays the lowest correlation with the market at -0.21 with the second lowest being the \text{GAT}$_\text{industries}$ with -0.11. Moreover, the original GAT's returns and signal is most strongly correlated with the \text{GAT}$_\text{corr}$'s returns and signal at 0.65 and 0.28 respectively compared with 0.52 and 0.28 for the \text{GAT}$_\text{del\_edge}$. This suggests the signal of the \text{GAT}$_\text{analysts}$ is reasonably different to other setups. The addition of attention has a significant effect on the signal as evidenced by the comparatively low correlations signal between the GCN and all of the other models which use the analyst matrix and especially the \text{GAT}$_\text{analysts}$.  This can be interpreted as an example of the differing effect of network information contained in each of the different networks upon the final trading signal: the analyst network helps the \text{GAT}$_\text{analysts}$ learn links between firms that are similar to lagged correlations and to a lesser extent industry linkages and yet distinct from both. 

%


\subsection{Turnover and cost analysis}
\label{subsec:turnover}

\begin{figure}[h!]
\centering
\begin{subfigure}[t]{0.5\linewidth}
    \centering
    \includegraphics[width=0.99\linewidth]{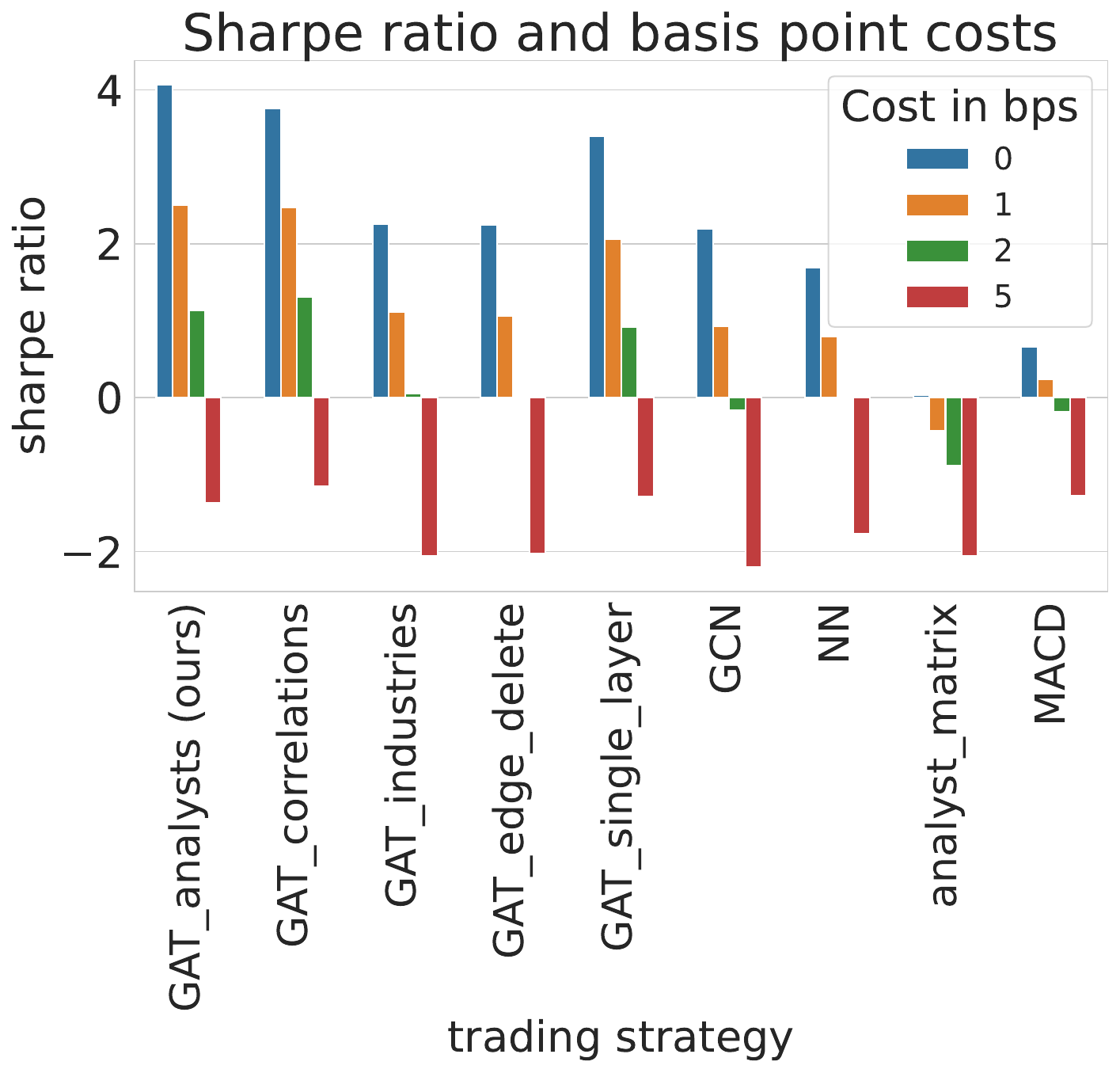}
    \caption{Trading cost and Sharpe}
    \label{fig:sharpe-decay}
\end{subfigure}
\begin{subfigure}[t]{0.5\linewidth}
    \centering
    \includegraphics[width=0.99\linewidth]{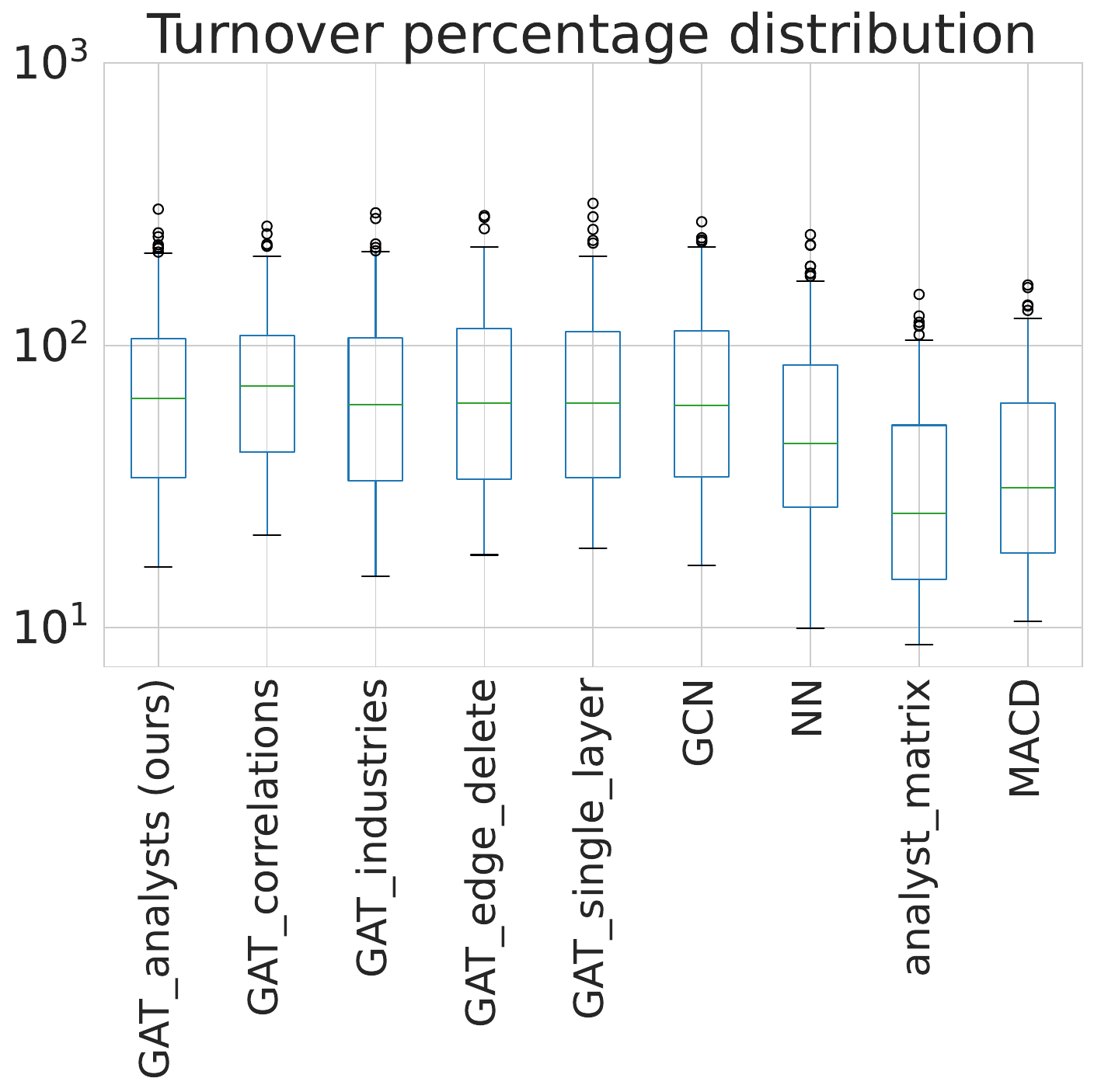}
    \caption{Turnover by strategy}
    \label{fig:turnover}
 \end{subfigure}
 \caption{Effect of trading cost on different strategies given turnovers}
\end{figure}




We investigate how the strategy performs when exposed to transaction costs in Figure \ref{fig:sharpe-decay}. This figure describes the annualized Sharpe ratio of the different strategies under increasing trading costs from 0 basis points increasing to 1, 2 and 5 basis points. This penalises strategies with larger turnovers in portfolio content as depicted in Figure \ref{fig:turnover}. We observe that the returns of all of the models under consideration a sizeable decay in performance as the trading costs increase. All of the strategies yield negative Sharpe ratios when exposed to 5 basis points of trading costs.  The \text{GAT}$_\text{analysts}$ and \text{GAT}$_\text{corr}$ display similar turnover than other strategies with comparable Sharpe ratios such as \text{GAT}$_\text{1\_layer}$. Only \text{GAT}$_\text{analysts}$, \text{GAT}$_\text{corr}$ and \text{GAT}$_\text{1\_layer}$ maintain positive Sharpe ratios when exposed to 2 basis points of trading costs: every other strategy exhibits a null or negative Sharpe ratio. This suggests that the analyst and correlation based strategies are more robust to trading frictions despite the drop in performance but it is the initial strong-risk adjusted performance that ensures they do well under tthese frictions. The \text{GAT}$_\text{del\_edge}$, \text{GAT}$_\text{industries}$ and $\text{GCN}$ display comparable turnover ratios. 
The analyst matrix's Sharpe ratio decays quickly from 0.069 to -2, however since it displays relatively less turnover, it outperforms certain more complex strategies such as $\text{GCN}$ or \text{GAT}$_\text{industries}$ under the highest cost regime. Lastly, we note that the Sharpe ratio of the MACD strategy fares the best out of all of the when exposed to the highest level of transaction costs, it experiences the least sharp drop going from 0.66 to -1.27. However, we also note that the \text{GAT}$_\text{analysts}$ approach still displays a better Sharpe ratio at a 2 basis point cost than the 0 transaction cost MACD portfolio and still outperforms it at the 5 basis points trading scenario. This suggests that the model-free approach, though interesting in terms of turnover-limitation, can be substituted under most cost scenarios with the proposed solution of this paper. Model-based approaches such as the NN, the \text{GAT}$_\text{analysts}$ and the ablated GAT models have on average a slightly higher average turnover (77\$) than the model-free approaches (40\$) such as the MACD and the analyst-matrix confirming findings in the literature \cite{lim2019enhancing}. Consequently, this explains why these model-free strategies are less affected by the increase in trading costs.




\subsection{Attention analysis}
\label{subsec:attention}

\begin{figure}[h!]
    \centering
    \includegraphics[width=\linewidth]{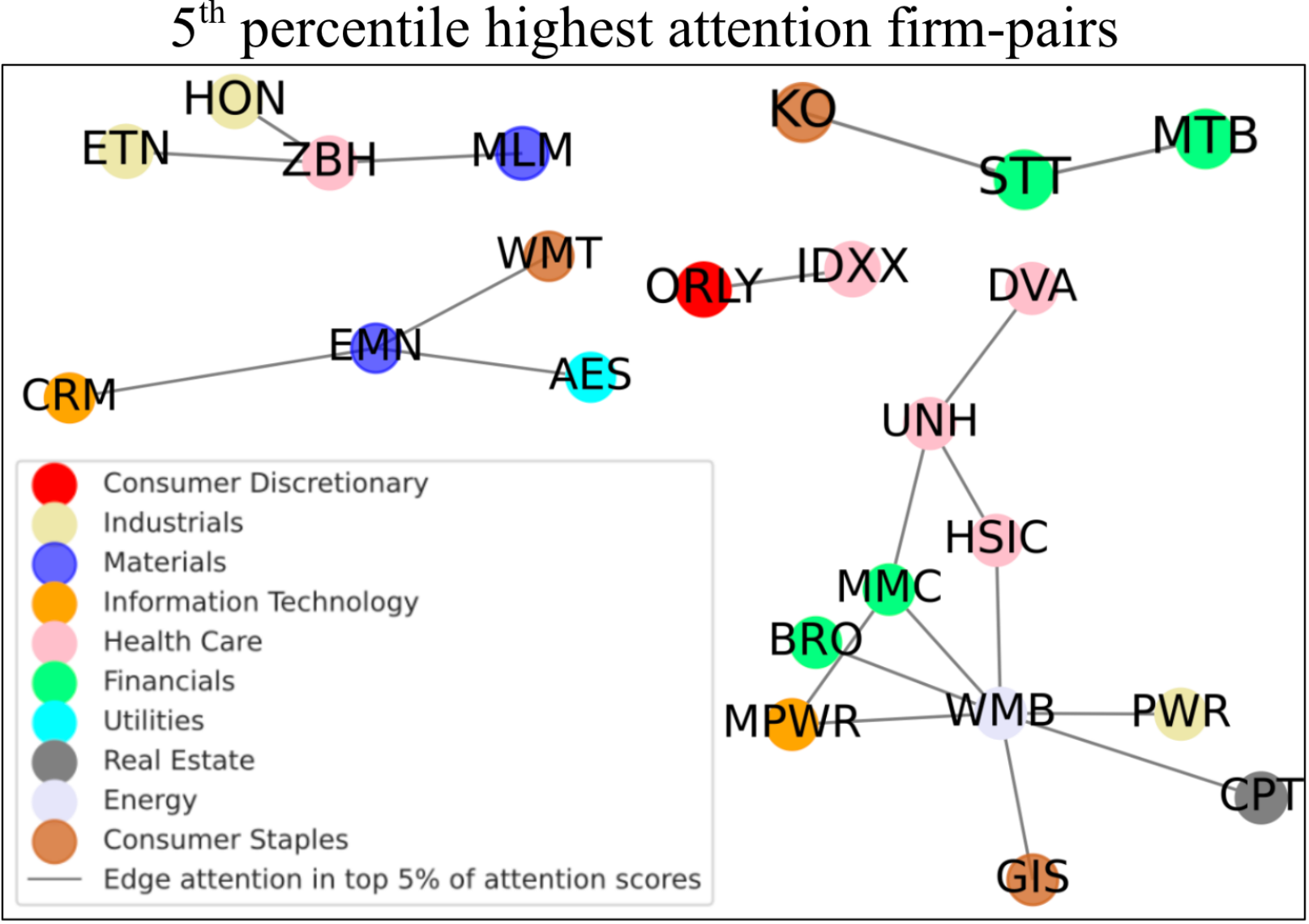}
    \caption{Firm links with strongest attention}
    \label{fig:topattention}
\end{figure}


Figure \ref{fig:topattention} represents the subset of the analyst matrix where the edges had the strongest attention coefficient in the GAT$_\text{analysts}$ model for one trading period in December 2016. The GAT$_\text{analysts}$ model learns to place high weight on the relationship between those firms when determining how to best forecast future value. The firm nodes are coloured by industry.  This figure reinforces the idea that the attention function is helpful in allowing the model to learn informative and interpretable links between firms .  For instance, the KO (Coca-cola) - STT (State Street Corp) link in the middle corresponds to a link between an investment firm (STT) with a very high stake in Coca-Cola. Similarly, the WMB cluster on the bottom right corresponds to an energy provider (WMB) connected to several firms who require energy provision, from the agro-industrial company GIS, to healthcare firms like HSIC. These links support the hypothesis that the GAT model identifies links between firms which correspond to fundamental economic linkages likely to lead to momentum spillover effects from investor under-reaction \citep{ali2020shared,lee2017uncovering,bekkerman2023effect}. 
Moreover, this graph shows the strongest attention weights being quite spread out by industry, i.e., the frequency of connection between industries is higher than within industries, suggesting that the model is able to flexibly learn inter and intra-industry patterns in a richer way than just focusing on the industry matrix as suggested by \cite{martens2021analyst}. 
Lastly, one important point to note is that none of these entries with highest attention are present in the correlation matrix as the strength of these correlations is lower than the 90-th percentile of correlation used in defining the correlation matrix. This implies that the attention mechanism applied to the analyst matrix is flexible enough to allow the model to reveal meaningful `economic relations' between firms \citep{cohen2008economic}. These links go beyond simple correlation based measures which have been criticized in the literature for an inability to capture non-linear relationships \citep{marti2021review}.

\section{Conclusion}
In this paper, we have explored the ability of a graph attention network to systematically learn a novel trading signal from a metric of firm to firm momentum spillovers: the analyst coverage network. We have shown that the information contained in the analyst network can be efficiently extracted by the graph attention network in order to produce positive and persistent out-of-sample trading returns. Our method outperforms existing benchmarks in terms of average returns and drawdowns. Furthermore, we have shown the strategy's robustness through ablation studies and turnover analysis. Moreover, we have shown the link between topological information contained in the analyst network and the performance of our strategy by comparing the performance of the model trained on the analyst network to that of the same model trained with industry and correlation based networks. This work represents the first step in incorporating analyst coverage networks into financial graph machine learning. We demonstrate the profitability of the strategy and the wealth of insights which can be leveraged from this setup. We outline several pathways for future work. The first is extending the proposed model to incorporate richer temporal information, for instance by making the edge information into a time series whose features could be modelled explicitly by a dedicated temporal learning block. Moreover, given the link between analyst estimation error and volatility, the GAT model could be used to model volatility spillover effects between firms across edges of the analyst network.

\bibliographystyle{ACM-Reference-Format}
\bibliography{references}

\end{document}